\documentclass[12pt]{article} 

\usepackage[utf8]{inputenc} 

\usepackage{geometry} 
\usepackage{graphicx} 

\usepackage[utf8]{inputenc} 
\usepackage{etex}
\usepackage{graphicx}
\usepackage{dcolumn}
\usepackage{amssymb,amsthm,amscd,amsbsy,array}
\usepackage{amstext}
\usepackage{amsmath,dsfont}
\usepackage{soul} 
\usepackage{graphics,xcolor}
\usepackage{bm}
\usepackage{booktabs} 
\usepackage{array} 
\usepackage{paralist} 
\usepackage{verbatim} 
\usepackage{subfig} 

\numberwithin{equation}{section}

\let\ssection=\section
\renewcommand{\section}{\setcounter{equation}{0}\ssection}
\textheight=
250mm 
\newcommand{\half}{{\frac{1}{2}}}
\def\2{{\half}}

\def\beq{\begin{equation}}
\def\eeq{\end{equation}}
\def\beqa{\begin{eqnarray}}
\def\eeqa{\end{eqnarray}}

\def\barray{\left(\begin{array}}
\def\earray{\end{array}\right)}
\def\barraynb{\begin{array}}
\def\earraynb{\end{array}}

\def\smallover#1/#2{\hbox{$\textstyle\frac{#1}{#2}$}} %

\newcommand{\cQ}{\mathcal{Q}}

\usepackage{color}

\newcommand{\gb}{\colorbox{green}}

\def\benu{\begin{enumerate}}
\def\eenu{\end{enumerate}}

\def\?{{\;\gb{\,?\,} \;}}

\def\cpp{{circularly polarized periodic gravitational wave\,}}
\def\PUo{{Pais-Uhlenbeck oscillator\,}}

\usepackage{fancyhdr} 
\usepackage{sectsty}
\allsectionsfont{\sffamily\mdseries\upshape} 

\usepackage[nottoc,notlof,notlot]{tocbibind} 
\usepackage[titles,subfigure]{tocloft} 

\title{Circularly polarized periodic gravitational wave and the Pais-Uhlenbeck oscillator }
\author{Mahmut Elbistan\footnote{mailto: mahmut.elbistan@boun.edu.tr}}

\date{} 

\begin{document}
\maketitle

\qquad \qquad Physics Department,
Bo\u{g}azi\c{c}i University,
34342 Bebek / Istanbul, Turkey

\vspace{1cm}

\begin{abstract}

The ``Bargmann framework'' allows us to prove the equivalence of
the Pais-Uhlenbeck oscillator with a spin-zero particle in a circularly polarized periodic gravitational wave. Pushing forward the symmetries (which include Carroll symmetry) of the latter, we derive conserved charges for the Pais-Uhlenbeck oscillator.

\end{abstract}

\tableofcontents

\pagebreak

\section{Introduction}

Recently in \cite{ilderton}, it was reported that a \emph{circularly polarized periodic gravitational wave} (CPP GW) \cite{phonebook} is the classical double copy of the electromagnetic vortex \cite{BBvort}. Soon after this, in \cite{IONGW}, periodic GWs were linked to ion traps \cite{Paul, Dehmelt} by Eisenhart-Duval (ED) lift aka "Bargmann framework" \cite{Eisenhart, DuvalBargmann, DGH}. 

In this note, firstly we reveal a novel correspondence between the 4-dimensional CPP GW and a $1$-dimensional, $4$th order \emph{Pais-Uhlenbeck (PU) oscillator} \cite{PU} (for a recent review see \cite{smilga}) via ED lift. More precisely, we show that the underlying non-relativistic (NR) dynamics of a CPP GW can be mapped to that of PU oscillator and vice versa. 

Generic isometries of plane GWs were identified as the 2+1 dimensional Carroll group with broken rotations \cite{Sou73, Carroll4GW}. Carroll group was derived as a novel contraction of Poincar\'e group a long time ago \cite{LL1, SenG} and has attracted a lot of recent attention \cite{ConfCarroll, Carrollvs, NewCarroll, Morand:2018tke,Ciambelli:2019lap,Marsot:2021tvq}. 
In addition to that 5-parameter Carroll symmetry, CPP GW is endowed with an extra ``screw" symmetry \cite{phonebook} and a homothety (see \cite{homoGW} for a recent treatment). 

Having established the correspondence of the CPP GW and the PU oscillator, secondly we push forward all those symmetries of the former and derive conserved quantities for the PU oscillator with arbitrary frequencies. In particular, Carroll translations and Carroll boosts generate spatial translation, boost and two accelerations for the PU oscillator. 
It was found before that the symmetries of the PU oscillator and its generalization span the conformal Newton-Hooke group provided that oscillator frequencies follow an arithmetic sequence \cite{GaMa1, AGGM}. It is also known that the symmetry algebra is simpler for generic frequencies \cite{Andr0} and our findings fit this scheme.

In \cite{GaMa2}, Galajinsky and Masterov studied the ED lift of higher derivative theories in a systematic way and illustrated it by the PU oscillator. However, their metric is not that of a plane-wave. Therefore, the link with the CPP GW and its symmetries is not apparent. 

A canonical transformation maps the hamiltonian of the PU oscillator into a pair of uncoupled oscillators with a relative minus sign \cite{smilga}. One can easily lift this system to higher dimension, -- but the resulting metric is non-Lorentzian.
A further transformation called the chiral decomposition \cite{Pl1, Pl2, Kohn} allows us to link to CPP GW and to pull back its conserved quantities.

In the next section, we discuss symmetry properties of the CPP GW. In Section 3, we show that projected dynamics is equivalent to that of a 1 dimensional PU oscillator. In Section 4, we derive conserved charges for the PU oscillator using symmetries of the CPP GW. We discuss our results in Section 5.

\section{Circularly polarized periodic gravitational wave in Brinkmann coordinates}

\subsection{Symmetries of the $4$-dimensional \cpp }

We first recall some properties of CPP GW that we need for our investigations (see \cite{polper} for more details). In Brinkmann coordinates $\{X^i, U, V \} $ \cite{Brinkmann}, a 4-dimensional CPP GW is given by
\begin{eqnarray}
\label{CPP}
ds^2&=& d\bm{X}^2 + 2 dU dV +  \frac{A_0}{2} \Big( \cos(\omega U)  \big( (X^1)^2 - (X^2)^2  \big)  +2 \sin(\omega U) X^1 X^2  \Big)dU^2 \\
&\equiv&d\bm{X}^2 + 2 dU dV + K_{ij}(U) X^i X^j dU^2. \nonumber 
\end{eqnarray}
Its physical properties are determined by its amplitude $A_0 > 0$ and its frequency $\omega$.  
Being an exact plane GW, its generic isometry group is the $2+1$ dimensional Carroll group with broken rotations. The associated 5-parameter Killing vector is
\begin{eqnarray}
\label{CarrollB}
Y_C &=& h\frac{\partial}{ \partial V} + c^i \left( L_{ji}\frac{\partial}{\partial X^j} - L'_{ji} X^j\frac{\partial}{\partial V} \right) + b_i \left(L_{jk}S^{ki} \frac{\partial}{\partial X^j} - ( L_{jk}S^{ki})' X^j \frac{\partial}{\partial V}\right) \quad \nonumber \\
&\equiv&  hY^v  + c^i Y^c_i  + b^i Y_i^b ,
\end{eqnarray}
where $L(U)$ is a $2\times 2$ matrix which satisfies a Sturm-Liouville equation augmented with a subsidiary condition,
\beq
\label{SLP}
L_{ij}'' = K_{ik} L_{kj}, \quad  L^T L' = (L^T)' L ,    
\eeq
with $\{\,\}' = \frac{d}{dU}\{ \,\}$. 
The $2\times2$ matrix 
\beq
S^{ij}(U) = \int^U a^{ij}(\tilde{U})  d\tilde{U}
\label{Smatrix}
\eeq
 above is called the \emph{Souriau matrix} with $a^{ij} = (L^T L)^{-1}_{ij}$ is the inverse of $a=(a_{ij})$.

A CPP gravitational wave has, in addition, a ``screw'' (or helical) symmetry
\beq
\label{screw} 
Y_s = \partial_U + \frac{\omega}{2} \epsilon_{ij} X^i \partial_j,
\eeq
where $\epsilon_{ij}$ is the 2 dimensional Levi-Civita tensor.
Together with the homothetic vector field
\beq
\label{Yhom}
Y_h = 2 V \partial_V + X^i\partial_i, \quad   {\mathcal{L}}_{Y_h} g_{\mu\nu} = 2 g_{\mu\nu},
\eeq 
CPP GW enjoys a 7-parameter conformal algebra \cite{phonebook} with non-zero commutators such that
\beq
\label{4dalg}
[  Y^c_i, Y^b_j ] = - \delta_{ij} Y^v, \quad  [ Y_h, Y_C ] = - Y_C - h Y^v .
\eeq
In order to find out the commutator of the screw symmetry (\ref{screw}) with the Carroll vector field, one needs to solve the Sturm-Liouville problem (\ref{SLP}). We will deal with this issue later in Section \ref{conservedPU}. 

Conserved charges regarding (\ref{CarrollB}) are given as
\begin{eqnarray}
\label{CarrollBcharges}
\cQ_C &=& h P_V + c^i \left( P_j L_{ji} - P_V X^j L'_{ji}  \right) + b_i \left( P_j  L_{jk}S^{ki} - P_V X^j ( L_{jk}S^{ki})'  \right) \quad \nonumber\\
&\equiv&  h \cQ^v  + c^i \cQ^c_i  + b^i \cQ_i^b, 
\end{eqnarray}
with ${P_\mu }$ being the canonical momenta of a spinless test particle in this background. Note that the charges $\cQ^c_i$ and $\cQ^b_i$ are related with the interchange $L \leftrightarrow LS $ \cite{zhangSL}. Being an isometry, $Y_s$ also yields a conserved quantity. On the other hand, the homothety (\ref{Yhom}) provides a conserved charge in the Noetherian sense only for null geodesics.


\subsection{Dynamics in rotating Brinkmann coordinates}

The geodesic equations of the CPP metric (\ref{CPP}) are complicated and to solve them we switch to rotating coordinates  $Y^i = (R^{-1})_{ij} X^j$ 
\begin{equation}
\label{introrot}
\begin{pmatrix}
 Y^1 \\ Y^2
\end{pmatrix}
= \begin{pmatrix}
\cos(\omega U/2)  &  \sin(\omega U/2) \\
-\sin(\omega U/2)  & \cos(\omega U/2)\,
\end{pmatrix}
\begin{pmatrix}
 X^1 \\ X^2
\end{pmatrix}\,,
\end{equation}
where $U$ and $V$ are kept fixed. Then (\ref{CPP}) becomes
\beq
\label{rotCPP}
ds^2 = d \bm{Y}^2 + 2 dU dV + \omega \epsilon_{ij} Y^i dY^j dU + \big(\Omega_+^2 (Y^1)^2 + \Omega_-^2 (Y^2)^2  \big ) dU^2,
\eeq
with
\beq
\label{Omegarel}
\Omega_\pm^2 = \frac{\omega^2}{4} \pm \frac{A_0}{2}. 
\eeq
In rotating coordinates, the Carroll symmetries (\ref{CarrollB}) are 
\begin{eqnarray}
\label{CarrollrotB}
Y_C &=& h\frac{\partial}{ \partial V} + c^i \left( L_{ji} R^{jk}\frac{\partial}{\partial Y^k} - L'_{ji} R^{jk} Y^k\frac{\partial}{\partial V} \right) + b_i \left(L_{jk}S^{ki} R^{jm} \frac{\partial}{\partial Y^m} - ( L_{jk}S^{ki})'  R^{jm} Y^m \frac{\partial}{\partial V}\right), \quad \quad 
\end{eqnarray}
where $R$ is the inverse of the matrix in (\ref{introrot}). 

Beginning with an affinely parametrized geodesic Lagrangian $L^{rot}_{4d} = \frac{1}{2} g_{\mu\nu} \dot{Y}^\mu \dot{Y}^\nu$ of the test particle with coordinates $Y^{\mu} = \{ Y^i, U, V  \}$, we obtain its Hamiltonian as
\beq
\label{H4Drot}
H^{rot}_{4d} = \frac{(P_i + \frac{\omega}{2} \epsilon_{ij} Y^j P_V)^2}{2} + P_U P_V - \frac{1}{2} \big( \Omega_+^2 (Y^1)^2  + \Omega_-^2 (Y^2)^2 \big)P_V^2 , 
\eeq 
where the canonical momenta $P_{\mu}$ are
\begin{eqnarray}
\label{H4Dmom}
P_i = \dot{Y}^i - \frac{\omega}{2} \epsilon^{ij} Y^j \dot{U},    \quad P_U = \dot{V} + \frac{\omega}{2} \epsilon^{ij} Y^i \dot{Y}^j + \big( \Omega_+^2 (Y^1)^2 + \Omega_-^2 (Y^2)^2 \big) \dot{U}, \quad P_V = \dot{U}. \quad
\end{eqnarray}
The new metric in rotating coordinates (\ref{rotCPP}) is $V$-independent, as is the original Brinkmann metric (\ref{CPP}). Thus, $P_V$ is a constant of the motion.  It is one of the conserved Carroll charges $\cQ_C$ 
\beq
\label{conrel}
\dot{\cQ}_C  = \{\cQ_C,  H^{rot}_{4d} \} = 0,
\eeq
obtained from the Killing vector (\ref{CarrollrotB}) via $\partial_\mu \to P_\mu$.  Here $\dot{()}$ is derivation w.r.t. an arbitrary affine parameter. Henceforth, we will choose it as $U$.
 
When we push forward $Y_s$ to rotating coordinates,  it becomes
\beq
\label{screwrot}
Y_s = \partial_U,
\eeq  
implying the $U$-independence of (\ref{rotCPP}). Thus, $P_U$ is conserved in rotating coordinates. This fact will allow us to solve the transverse geodesic equations in Section (\ref{chiraldecomposition}).

\section{Projecting the relativistic dynamics to the \PUo} 

Now, using the Bargmann language  \cite{Eisenhart, DuvalBargmann, DGH}, we project the relativistic dynamics in the GW spacetime to the underlying non-relativistic space-time. 
Considering massless geodesics and thus setting $H^{rot}_{4D}$ in (\ref{H4Drot}) to vanish, 
\beq
H^{rot}_{4d} = \frac{g^{\mu\nu}_{rot} P_\mu P_\nu}{2}\equiv 0 , 
\eeq
we obtain a NR Hamiltonian 
\beq
\label{HNRrot}
H^{rot}_{NR} = \frac{ (P_i + \frac{M\omega}{2}  \epsilon_{ij} Y^j )^2  }{2M}  - \frac{M}{2} \big( \Omega_+^2 (Y^1)^2  + \Omega_-^2 (Y^2)^2 \big) = - P_U.
\eeq
The light-cone coordinate $U$ becomes Newtonian time and $P_V = M$ becomes the mass of the NR particle. Hamilton equations are found to be
\beq
\label{NRHD}
(Y^i)' = \frac{P_i + \frac{M\omega}{2}  \epsilon_{ij} Y^j }{M} , \quad  (P_i)' = \frac{\omega}{2} \epsilon_{ij} (P_j + \frac{M\omega}{2}  \epsilon_{jk} Y^k ) + M ( \Omega_+^2 Y^1 \delta_{i, 1}  + \Omega_-^2 Y^2 \delta_{i,2} ). 
\eeq
The dynamics generated by $H^{rot}_{NR}$ (\ref{HNRrot}) is a combination of attractive and repulsive oscillators plus a Coriolis force \cite{IONGW}. Note that $H^{rot}_{NR}$ does not explicitly depend on $U$ i.e., the total energy is conserved as implied by the screw symmetry ({\ref{screwrot}}).

From (\ref{NRHD}) we obtain two coupled, second order Euler - Lagrange equations 
\begin{subequations}
\label{Yeqnsrot}
\begin{align}
(Y^1)'' - \omega (Y^2)' - \Omega_+^2 Y^1 &= 0 \\
(Y^2)'' + \omega (Y^1)' - \Omega_-^2 Y^2 &= 0,
\end{align}
\end{subequations}
with \emph{time-$U$ independent} coefficients. 

Before solving them, let us make an important observation. Equations (\ref{Yeqnsrot}) can be cast into the form of a \emph{4th order differential equation} for either of the components~: 
we take one of them, say (\ref{Yeqnsrot}a),  derive it with respect to $U$.  Substituting $(Y^2)''$ from (\ref{Yeqnsrot}b) and taking derivative once more we obtain a 4th order but coupled equation. Finally, substituting $(Y^2)'$ from (\ref{Yeqnsrot}a), we get a 4th order equation for $Y^1$ alone. Beginning instead with (\ref{Yeqnsrot}b) we would get another 4th order equation for $Y^2$. 
For either of them we get a higher derivative equation
\beq
\label{CPPU}
(Y^i)'''' + (\Omega_+^2 + \Omega_-^2) (Y^i)'' + \Omega_+^2 \Omega_-^2 Y^i =0,
\eeq
 which is exactly the equation of motion for the 1-dimensional, 4th order PU oscillator \cite{PU}. Note that both systems i.e., (\ref{Yeqnsrot}) and (\ref{CPPU}) have a 4-dimensional phase space.

Thus we conclude that, in rotating coordinates, the underlying NR system of a CPP GW can be described by the PU oscillator (\ref{CPPU}). 

Next we will analyze this result by relating  $H^{rot}_{NR}$ to the PU Hamiltonian. Henceforth, we will assume that $\Omega_-^2 >0$. This will allow us to compare our results with those of \cite{AGGM, Andr0} where real and non-negative frequencies are considered.
However, as it can be seen from (\ref{Omegarel}), $\Omega_-^2$ can also be negative\footnote{See the discussion about the complex frequencies in the novel paper \cite{PU}.} or vanishing and below we will comment on those cases too.

\subsection{Chiral decomposition }
\label{chiraldecomposition}

We first recall how the CPP equations (\ref{Yeqnsrot}) can be solved by chiral  decomposition \cite{Pl1, Pl2, Kohn}. The related symplectic structure\footnote{Written in terms of canonical coordinates, the symplectic 2-from becomes $\sigma = d P_1 \wedge dY^1 + dP_2 \wedge dY^2$ as expected.} for (\ref{Yeqnsrot}) is
\begin{eqnarray}
\label{sigmarot}
\sigma &=& d \Pi_1 \wedge dY^1 + d\Pi_2 \wedge dY^2 + \omega M dY^1 \wedge d Y^2, \\
H^{rot}_{NR} &=& \frac{\bm{\Pi}^2}{2M} - \frac{1}{2} M (\Omega_+^2  (Y^1)^2 + \Omega_-^2 (Y^2)^2), 
\label{Hpirot}
\end{eqnarray}
where the $\Pi_i = P_i +\frac{M \omega}{2} \epsilon_{ij} Y^j  $  are the kinematical momenta.  Chiral decomposition is based on finding ``good coordinates" such that our coupled system is decomposed into two disjoint parts.  One set of such coordinates  is $(X_+^{1,2}, \ X_-^{1,2} )$ such that
\begin{subequations}
\label{decomposec}
\begin{align}
\Pi_1 &= \frac{M\omega}{2} \left( X_+^2 + \frac{4\Omega_-^2}{\omega^2 } X_-^2 \right), \quad  \Pi_2 = - \frac{M\omega}{2} \left(  \frac{4 \Omega_+^2}{\omega^2}   X_+^1 + X_-^1 \right), \\
Y^1 &= X_+^1 + X_-^1, \quad  Y^2 = X_+^2 + X_-^2 .  
\end{align}
\end{subequations}
Then both the symplectic form $\sigma$ (\ref{sigmarot}) and the Hamiltonian $H^{rot}_{NR}$ (\ref{Hpirot}) are decomposed into separate $\pm$ sectors, 
\begin{eqnarray}
\label{dsigma}
\sigma &=& - \frac{M (\Omega_+^2 - \Omega_-^2)}{\sqrt{2 (\Omega_+^2 + \Omega_-^2)}}  (  dX_+^1 \wedge dX_+^2 - dX_-^1 \wedge dX_-^2  ) , \\
H^{rot}_{NR} &=& \frac{M (\Omega_+^2 - \Omega_-^2)}{2}  \left( \frac{\Omega_+^2}{\Omega_+^2 + \Omega_-^2} (X_+^1)^2 + \frac{1}{2} (X_+^2)^2   - \frac{1}{2}  (X_-^1)^2 -  \frac{\Omega_-^2}{\Omega_+^2 + \Omega_-^2} (X_-^2)^2        \right).  \qquad \qquad
\label{dHamilton}
\end{eqnarray}
We can easily find the equations of motion as
\beq
(X_\pm^{1,2} )'' + \Omega_\pm^2 X^{1,2}_\pm = 0 ,
\eeq
which are nothing but simple harmonic motions\footnote{We consider parameter values such that $\Omega_-^2 > 0$ to stay in that realm.} with frequencies $\Omega_\pm$.  The solutions for $Y^{1,2}$ in (\ref{decomposec}b) generalize the ones in  \cite{IONGW} for arbitrary $\omega$,\begin{subequations}
\begin{align}
Y^1(U)&= A_1 \cos (\Omega_+ U) + B_1 \sin (\Omega_+ U) +  C_1 \cos (\Omega_- U) + D_1 \sin (\Omega_- U) ,
\\
Y^2(U)&= - \Omega_+  \left(  \frac{2}{\Omega_+^2 + \Omega_-^2} \right)^{\frac{1}{2}}     \big( A_1\sin(\Omega_+ U) - B_1\cos(\Omega_+ U)\big) \nonumber \\
&-
\dfrac{1}{\Omega_-}  \left( \frac{\Omega_+^2 + \Omega_-^2}{2}  \right)^{\frac{1}{2}} \big( C_1\sin(\Omega_- U) - D_1 \cos (\Omega_- U)\big),
\end{align}
\label{polperLsolns}
\end{subequations}
where $A_1, B_1, C_1, D_1$ are constants to be determined from the initial conditions. $Y^{1,2}$ (\ref{polperLsolns}) also solve the 4th order PU equation (\ref{CPPU}). The geodesics of the original Brinkmann metric (\ref{CPP}) can be found by the inverse  transformation of (\ref{introrot}).

We emphasize that $\sigma$ (\ref{dsigma}) and $H^{rot}_{NR}$ (\ref{dHamilton})  never vanish since $\Omega_+ > \Omega_-$. By (\ref{Omegarel}), their equality  would imply zero wave amplitude $A_0 =0$ , i.e., flat Minkowski space-time. Apart from this, we do not require any other relation between them.

Following the same steps we can work out our system (\ref{sigmarot}) and (\ref{Hpirot}) for other frequency values. For a strong wave, i.e.,  $\Omega_-^2 < 0$, chiral decomposition results in a combination of an harmonic oscillator and an "inverted" oscillator. The latter reflects itself in (\ref{polperLsolns}) as an escaping solution in terms of hyperbolic functions with frequency $|\Omega_-|$. 
 On the other hand, if $\Omega_- =0$, then we again have an unbounded motion for the $-$ sector but this time in terms of a free particle. Explicit solutions for this case were already given in \cite{polper}, namely eqn. $\#$ (5.11) there.

\subsection{Explicit relation with the Pais-Uhlenbeck oscillator}

After this preparation, we return to the PU oscillator. The following redefinition 
\begin{subequations}
\label{scale}
\begin{align}
 x^1 &= \sqrt{\frac{M(\Omega_+^2 - \Omega_-^2)}{\Omega_+^2 + \Omega_-^2}} X_+^1 , \quad  p_1 = \sqrt{\frac{1}{2} M(\Omega_+^2 - \Omega_-^2)} X_+^2 , \\
 x^2 &= \sqrt{\frac{M(\Omega_+^2 - \Omega_-^2)}{\Omega_+^2 + \Omega_-^2}} X_-^2 , \quad  p_2 = \sqrt{\frac{1}{2} M(\Omega_+^2 - \Omega_-^2)} X_-^1 , 
\end{align}
\end{subequations}
allows us to pass to canonical coordinates $\{x^i, p_i\}$ such that (\ref{dHamilton}) and (\ref{dsigma}) become
\begin{eqnarray}
\label{Hopm}
H^{rot}_{NR}  \equiv H_{o\pm} =  \frac{p_1^2 + \Omega_+^2 (x^1)^2}{2} -  \frac{p_2^2 + \Omega_-^2 (x^2)^2}{2} ,   \qquad \sigma= dp_i \wedge dx^i .
\end{eqnarray}
$H_{o\pm} $ consists of two uncoupled harmonic oscillators but with a relative minus sign between them.  Thus, the related Hamilton equations are 
\beq
\label{Hpmeqns}
(x^1)' = p_1, \quad (p_1)' = - \Omega_+^2 x^1, \qquad  (x^2)' = - p_2, \quad (p_2)' = \Omega_-^2 x^2 .  
\eeq

A final canonical transformation \cite{smilga} 
\begin{subequations}
\label{scanon}
\begin{align}
x^1 &= \frac{ p_x +  \Omega_+^2 v }{\Omega_+ \sqrt{(\Omega_+^2 - \Omega_-^2)}},  \quad   p_1 = \frac{\Omega_+ (p_v + \Omega_-^2 x)}{ \sqrt{(\Omega_+^2 - \Omega_-^2)}}, \\
x^2 & = \frac{p_v + \Omega_+^2 x }{\sqrt{(\Omega_+^2 - \Omega_-^2)}} ,   \quad p_2  = \frac{p_x + \Omega_-^2 v}{ \sqrt{(\Omega_+^2 - \Omega_-^2)}},
\end{align}
\end{subequations}
yields the Ostrogradsky Hamiltonian for the PU oscillator,
\beq
\label{PUHorigin}
H_{o\pm}   = H_{PU} = p_x v + \frac{p_v^2}{2} + \frac{(\Omega_+^2 + \Omega_-^2)v^2}{2}  -  \frac{\Omega_+^2 \Omega_-^2 x^2}{2} .  
\eeq 
Therefore we have shown that, in addition to the equations of motion, the underlying NR Hamiltonian (\ref{HNRrot}) can also be mapped to the PU Hamiltonian (\ref{PUHorigin}).
Put differently: if we begin with PU Hamiltonian $H_{PU}$ (\ref{PUHorigin}) and use the inverse transformations of (\ref{scanon}),  (\ref{scale}) and (\ref{decomposec}) respectively, we end up with $H^{rot}_{NR}$ (\ref{HNRrot}).  Unrotating with inverse of (\ref{introrot}) and ED lifting it, we obtain the Brinkmann CPP metric (\ref{CPP}).
Thus we have found that the PU oscillator is the underlying NR system of a CPP GW \footnote{See \cite{Sarioglu:2006vc} for another realization of the PU oscillator.}.

To complete this part, we note the associated Lagrangian  
\beq
L_{PU} = \frac{1}{2} \Big( {x''}^2 - (\Omega_+^2 + \Omega_-^2){x'}^2 +\Omega_+^2\Omega_-^2 x^2 \Big),
\label{PULag}
\eeq
and the equation of motion 
 \beq
 \Big(\frac{d^4}{du^4} + (\Omega_+^2 + \Omega_-^2)\frac{d^2}{du^2} + \Omega_+^2\Omega_-^2 \Big) x=0,
 \label{PUeqns}
 \eeq
cf. (\ref{CPPU}) with $u =U$ and $\Omega_+ > \Omega_-$.

The PU oscillator phase space is $4$ dimensional with coordinates $\{x, p_x, v, p_v \}$ where $v = x'$. The definition of the canonical momenta ,
 \beq
 p_v = \frac{\partial L_{PU}}{\partial x''}, \quad p_x = \frac{\partial L_{PU}}{\partial x'} -\frac{d}{dt}  \frac{\partial L_{PU}}{\partial x''} =\frac{\partial L_{PU}}{\partial x'}- {p_v}',
 \eeq
 yields the Ostrogradsky Hamiltonian (\ref{PUHorigin}). The Hamilton equations of motion are 
\begin{subequations}
\label{PUHeqns}
\begin{align}
x' &\equiv \frac{\partial H_{PU}}{\partial p_x} = v,    \quad   v' \equiv \frac{\partial H_{PU}}{\partial p_v} = p_v , \\
{p_x}' &\equiv - \frac{\partial H_{PU}}{\partial x} = \Omega_+^2 \Omega_-^2 x, \quad  {p_v}' \equiv - \frac{\partial H_{PU}}{\partial v} = - p_x - (\Omega_+^2 + \Omega_-^2) v .
\end{align}
\end{subequations}

\section{Conserved charges of the \PUo}
\label{conservedPU}

Now we derive the conserved charges of the PU oscillator from those of a spinless test particle in a CPP GW background. Because of projection these charges will have an explicit time dependence,
\beq
\label{NRcon}
\frac{d \cQ}{dU} = \frac{\partial \cQ}{\partial U} + \{ \cQ, H_{NR} \}  = 0, 
\eeq
cf. (\ref{conrel}).
Note that direct projection of (\ref{CarrollBcharges}) to the underlying NR system yields the following Poisson algebra  
\beq
\label{CarrollBQ}
\{ \cQ_i^c, \cQ_j^b \} = \delta_{ij} M. 
\eeq

CPP GW (\ref{CPP}) has 7 symmetries. The isometries  are generated by the 5 dimensional Carroll algebra  (\ref{CarrollB}). In addition, we have the screw isometry (\ref{screw}) and a homothetic vector field (\ref{Yhom}). 

In the original Brinkmann coordinates (\ref{CPP}), the screw isometry yields the following conserved charge
\beq
\label{screwC}
\cQ_s = P_U + \frac{\omega}{2} \epsilon_{ij}X^i P_j\,,
\eeq
where $P_U$ and $P_i$ are the canonical momenta. 
In rotating coordinates, it becomes (\ref{screwrot})  and the related conserved charge is then $P_U$.  When projected, it corresponds to (minus) the \emph{total conserved energy}  of the NR system (\ref{HNRrot}). 
After chiral decomposition it becomes $\cQ_s =  -H_{o\pm}$ (\ref{Hopm}) . Finally, performing a second transformation (\ref{scanon}), it yields the conserved energy of the PU oscillator (\ref{PUHorigin}) up to a minus sign,  
\beq
\label{screwPU}
\cQ_s = - H_{PU}. 
\eeq

The homothetic vector field (\ref{Yhom}) does not change in rotating coordinates.
 When projected down, it becomes a conserved quantity \cite{homoGW}
\beq
\label{Qhomrot}
\cQ_h = Y^i P_i + 2 M V, \quad 
V(U) = V_0 - \frac{1}{M} \int^U L_{NR}\ d\tilde{U} ,
\eeq
with $L_{NR}$ being the associated NR Lagrangian
\beq
\label{LNR}
L_{NR}^{rot} = \frac{1}{2} M (\bm{Y}')^2 - \frac{M \omega}{2} \epsilon_{ij} (Y^i)' Y^j + \frac{M}{2} \big( \Omega_+^2 (Y^1)^2 + \Omega_-^2 (Y^2)^2)  
\eeq
and $V_0$ is a constant. For various physical systems the homothetic charge was shown to be related to the virial theorem \cite{homoGW, noether}. 

In oscillator coordinates $\{x^i, p_i\}$, (up to a overall minus sign)  ${\cal{Q}}_{h}$ turns out to be  
\beq
\label{Qhopm}
\cQ_h = p_i x^i + \sqrt{2(\Omega_+^2 + \Omega_-^2)} x^1 x^2 - \int^u  d\tilde{u} \ \Big(2 L_{o\pm} + \sqrt{2(\Omega_+^2 + \Omega_-^2)} (p_1 x^2 - p_2 x^1)\Big) .
\eeq
Its conservation can be easily confirmed by using (\ref{Hopm}).

As in oscillator coordinates the homothetic charge 
\beq
\cQ_h = \cQ_{h1} + \cQ_{h2}
\eeq
 again consists of two parts which are separately conserved in PU-phase space $\{ x, v, p_x, p_v \} $\footnote{A constant factor $\frac{\sqrt{\Omega_+^2 +\Omega_-^2} (\sqrt{2} \Omega_+ + \sqrt{\Omega_+^2 +\Omega_-^2} ) }{\Omega_-^2 - \Omega_+^2}$ is omitted. Remember that the denominator  does not vanish.}, 
\begin{eqnarray} 
\label{cQh1PU}
\cQ_{h1} &=& p_x x + p_v v - 2\int d \tilde{u}\ L_{PU}, \\ 
\cQ_{h2} &=&  p_x p_v + \frac{\Omega_+^3\big( \sqrt{2} \Omega_-^2 + \Omega_+ (\Omega_+^2 + \Omega_-^2)^{1/2}   \big)}{(\Omega_+^2 + \Omega_-^2)^{1/2} + \sqrt{2} \Omega_+ }  xv   \nonumber \\
&-& \int^u d \left( p_x p_v + \frac{\Omega_+^3\big( \sqrt{2} \Omega_-^2 + \Omega_+ (\Omega_+^2 + \Omega_-^2)^{1/2}   \big)}{(\Omega_+^2 + \Omega_-^2)^{1/2} + \sqrt{2} \Omega_+ }  xv   \right) \qquad \qquad \qquad 
\label{CQh2PU}
\end{eqnarray}
While $\cQ_{h2}=0$ is trivial, the virial charge $\cQ_{h1}$ (\ref{cQh1PU}) originates from the scaling symmetry of the PU action (\ref{PULag})
\beq
x \to \lambda x,  \quad  L_{PU}\ dt \to \lambda^2 L_{PU}\ dt ,
\eeq
where $\lambda$ is a constant scale factor.

Now we arrive at the most important part. We will derive charges using the Carroll symmetry of the CPP GW.
In rotating coordinates (\ref{rotCPP}), the S-L problem (\ref{SLP}) is modified as 
\beq
\label{rotSLP}
O''_{ij} = \omega \epsilon_{ik} O'_{kj} + \Omega_{ik} O_{kj}, 
\eeq
where
\beq
\Omega = \begin{pmatrix} 
\Omega_+^2 & 0 \\
0 & \Omega_-^2 
\end{pmatrix} .
\eeq 
Above $2\times 2$ $O(U)$ matrix \cite{IONGW} is defined via the rotation in (\ref{introrot}) 
\beq
\label{Omatrix}
O_{ij} = (R^{-1})_{ik} L_{kj}.
\eeq
On the other hand, the subsidiary condition $ L^T L' = (L^T)' L$ turns out to be
\beq
\label{rotSLPs}
 \big(O^T O' - (O^T)' O\big)_{ij} = \omega O^T_{ik} \epsilon_{km} O_{mj},
\eeq 
which, in explicit form, becomes
\beq
O_{11} O'_{12} - O_{12} O'_{11} + O_{21}Q'_{22} - O_{22}O'_{21}   = \omega  \det{O} .
\eeq
Using the $O$-matrix (\ref{Omatrix}), the projected Carroll charges in rotating coordinates become
\begin{eqnarray}
\cQ_C &=& h M + c^i \left( P_k O_{ki}  - M Y^k ( O'_{ki}  - \frac{\omega}{2} \epsilon_{kj} O_{ji})  \right) \nonumber \\
&+&b_i \left( P_j O_{jk} S^{ki} - M Y^j ( O'_{jk} S^{ki} - \frac{\omega}{2} \epsilon_{jm} O_{mk} S^{ki} + O^{-1}_{ij})     \right) \nonumber \\
&\equiv& h \cQ_V + c^i \cQ_i^c + b^i \cQ_i^b .
\label{Carrollrotgen}
\end{eqnarray}

We note that (\ref{rotSLP}) has the same structure with (\ref{Yeqnsrot}). Thus, we may derive $O_{ij}$ by using the solutions (\ref{polperLsolns}) with 
\begin{equation}
O =
\begin{pmatrix}
Y^1 & \tilde{Y}^1 \\
Y^2 & \tilde{Y}^2
\end{pmatrix}
\label{Ogeneric}
\end{equation}
where $\tilde{Y}^i$ are of the same form as $Y^i$ but with different coefficients $A_2, B_2, C_2, D_2 $.

Let's discuss $\cQ_C$ (\ref{Carrollrotgen}). $P_V$  becomes the NR mass, i.e., $\cQ_V = M$,  a constant in PU phase space.

In oscillator coordinates $\{x^i, p_i  \}$ the projected translations and boosts $\cQ_i^c, \ \cQ_i^b$ (\ref{Carrollrotgen}) can be 
combined in the following form\footnote{We omit a constant factor $-\sqrt{\frac{M}{2(\Omega_+^2 - \Omega_-^2 )}}$.}
\begin{eqnarray}
\cQ_i^{c, b} &=& \Big(  \sqrt{2 (\Omega_+^2 + \Omega_-^2)} A_{1i}' - 2 \Omega_-^2 A_{2i} \Big) x^1 + \Big(  \sqrt{2 (\Omega_+^2 + \Omega_-^2)} A_{1i} + 2 A_{2i}' \Big) p_1 \nonumber \\
&+& \Big(  \sqrt{2 (\Omega_+^2 + \Omega_-^2)} A_{2i}' + 2 \Omega_+^2 A_{1i} \Big) x^2 -  \Big(  \sqrt{2 (\Omega_+^2 + \Omega_-^2)} A_{2i} - 2 A_{1i}' \Big) p_2
\label{QCpm}
\end{eqnarray}
where 
\begin{eqnarray}
A_{ji} &=& O_{ji},   \qquad \quad   A_{ji}' = O_{ji}'  \qquad \qquad \qquad \quad \text{for} \quad \cQ_i^c, \\
A_{ji} &=& O_{jk}S^{ki}, \quad A_{ji}' = O_{jk}'S^{ki} + O_{ij}^{-1} \quad \quad \quad \text{for} \quad \cQ_i^b .
\end{eqnarray}
Their conservation (\ref{NRcon}) can be proven via the Hamilton equations (\ref{Hpmeqns}) augmented with (\ref{rotSLP}) and (\ref{rotSLPs}). 

In PU coordinates $\{x, v, p_x, p_v \}$, omitting a constant factor  $-\sqrt{\frac{M}{2(\Omega_+^2 - \Omega_-^2 )^2}}$,  $\cQ_i^c$ and  $\cQ_i^b$ (\ref{QCpm}) become
\begin{eqnarray}
\cQ_i^{c, b} &=& \Omega_+ \left(  \big(2 \Omega_+^3 + \sqrt{2 (\Omega_+^2 + \Omega_-^2)} \Omega_-^2\big) A_{1i} + \big(2 \Omega_-^2 + \sqrt{2 (\Omega_+^2 + \Omega_-^2)} \Omega_+\big) A_{2i}'  \right) x \nonumber \\
&-&\frac{1}{\Omega_+} \left( \big(2 \Omega_-^2 + \sqrt{2 (\Omega_+^2 + \Omega_-^2)} \Omega_+\big) A_{2i} - \big(2 \Omega_+  + \sqrt{2 (\Omega_+^2 + \Omega_-^2)} \big) A_{1i}'     \right) p_x \nonumber \\
&-& \left(  \big(2 \Omega_-^2\Omega_+  + \sqrt{2 (\Omega_+^2 + \Omega_-^2)} \Omega_-^2\big) A_{2i}  - \big(2 \Omega_-^2 + \sqrt{2 (\Omega_+^2 + \Omega_-^2)} \Omega_+\big) A_{1i}'   \right) v \nonumber \\
&+& \left( \big(  2 \Omega_+  + \sqrt{2 (\Omega_+^2 + \Omega_-^2)} \big) ( \Omega_+ A_{1i} + A_{2i}'  ) \right) p_v .
\label{QCPU}
\end{eqnarray}
Again, they are conserved via the Hamilton equations (\ref{PUHeqns}) when (\ref{rotSLP}) and (\ref{rotSLPs}) are satisfied. However, this generic form is complicated if the  $O$-matrix (\ref{Ogeneric}) is directly substituted. Therefore, we simplify it by using the initial condition  
\beq
(X^i)' (0) = 0, 
\eeq
of the original 4 dimensional problem (\ref{Yeqnsrot}). This initial condition implies 
\beq
O (0) = 1_{2\times 2}, \quad  O'(0) = \frac{\omega}{2} 
\label{simpinitial}
\begin{pmatrix}
0 & 1 \\
-1 & 0
\end{pmatrix}, 
\eeq
which allows us to eliminate coefficients of the original complicated $O$-matrix\footnote{This generalizes the result in \cite{IONGW} for an arbitrary $\omega$. }
\beq
O = 
\begin{pmatrix}
\cos(\Omega_- u)  &  \frac{\omega}{2\Omega_+} \sin(\Omega_+ u) \\
-\frac{\omega}{2\Omega_-} \sin(\Omega_- u)  &  \cos(\Omega_+ u)   
\end{pmatrix},  \quad   \omega = \sqrt{2 (\Omega_+^2 + \Omega_-^2)}.
\label{Oshrunk}
\eeq
Then the Souriau matrix is obtained as
\beq
\label{Sshrunk}
S = \frac{2}{(\Omega_+^2 - \Omega_-^2) \det{O}} 
\begin{pmatrix}
\scriptstyle \frac{\omega^2}{4 \Omega_+} \cos(\Omega_- u) \sin(\Omega_+ u) - \Omega_- \sin(\Omega_- u) \cos(\Omega_+ u) & -\frac{\omega}{2} \\
\scriptstyle  -\frac{\omega}{2}  &\scriptstyle  \Omega_+ \cos(\Omega_- u ) \sin(\Omega_+ u) - \frac{\omega^2}{4 \Omega_-} \sin(\Omega_- u) \cos(\Omega_+ u) .
\end{pmatrix}
\eeq

Substituting (\ref{Oshrunk}) and (\ref{Sshrunk}) into (\ref{QCPU}) and replacing previously omitted coefficients, the projected Carroll charges become 
\begin{subequations} 
\label{QCarrollPU}
\begin{align}
\cQ_1^{c} &= - \sqrt{\frac{M}{2}}\left[\cos(\Omega_- u) \Big(\Omega_+^2 x + p_v\Big) + \sin(\Omega_- u) \Big(\frac{p_x}{\Omega_-} + \Omega_- v\Big) \right], \\
\cQ_2^{c} & = - \sqrt{\frac{M}{2}} \left[ \cos(\Omega_+ u) \Big(\frac{p_x}{\Omega_+} + \Omega_+ v\Big) - \sin(\Omega_+ u) \Big(\Omega_-^2 x + p_v   \Big) \right], \\ 
\cQ_1^b &= - \frac{\sqrt{2M}}{(\Omega_+^2 - \Omega_-^2)}\left[\cos(\Omega_- u) \Big(\Omega_-^2 v + p_x\Big) - \Omega_- \sin(\Omega_- u) \Big( p_v + \Omega_+^2 x   \Big) \right], \\
\cQ_2^b &=  - \frac{\sqrt{2M}}{(\Omega_+^2 - \Omega_-^2)}\left[\Omega_+ \cos(\Omega_+ u) \Big( \Omega_-^2 x + p_v  \Big) + \sin(\Omega_+ u) \Big(   \Omega_+^2 v + p_x  \Big) \right].
\end{align}
\end{subequations}
We emphasize that $( \cQ_i^c, \cQ_i^b )$ are conserved in PU phase space (\ref{PUHeqns}), i.e., 
\beq
\frac{d \cQ}{du} = \frac{\partial \cQ}{\partial u} + \{\cQ, H_{PU} \} = 0.
\eeq  

In PU phase space, the relation between  $\cQ_i^c$ and $\cQ_i^b$ becomes similar to EM duality (rotation) augmented with frequencies. For instance $\cQ_1^c$ and $\cQ_1^b$ are related via
\beq
\cos(\Omega_-u) \to - \frac{2\Omega_-}{(\Omega_+^2 - \Omega_+^2 )} \sin(\Omega_-u), \quad   \quad \sin(\Omega_- u) \to  \frac{2\Omega_-}{(\Omega_+^2 - \Omega_+^2 )} \cos(\Omega_-u),
\eeq  
cf. (\ref{CarrollBcharges}).
A similar relation holds between $\cQ_2^c$ and $\cQ_2^b$ with $\Omega_+$ and inverse rotation.

Non-vanishing Poisson brackets of $M$, $\cQ_i^c$ and $\cQ_i^b$ (\ref{QCarrollPU})
\beq
\label{QCarrollPUPoisson}
\{\cQ_i^c , \cQ_j^b \} =  M\delta_{ij},
\eeq
 span the algebra of the original but projected charges (\ref{CarrollBQ}) of the 4 dimensional CPP GW. To sum up, we have derived conserved charges of the PU oscillator from the Carroll symmetry of the plane GW.

Lastly, Poisson brackets with the screw charge $\cQ_s$ are found to be
\beq
\label{QCarrollHPUPoisson}
\{\cQ_{1,2}^c \ , H_{PU} \} = \mp\frac{(\Omega_+^2 - \Omega_-^2)}{2} \cQ_{1,2}^b\,, \quad    
\{\cQ_{1,2}^b\ , H_{PU} \} = \pm \frac{2\Omega_\mp^2}{(\Omega_+^2 - \Omega_-^2)}\cQ_{1,2}^c\, .
\eeq
Charges (\ref{screwPU}) and (\ref{QCarrollPU}) are in line with the ones found in \cite{Andr0} up to constant factors. Thus, Poisson brackets (\ref{QCarrollPUPoisson}) and (\ref{QCarrollHPUPoisson}) reproduce the same algebra for generic frequencies, namely eqns. $\#$ (2.5) and $\#$ (2.7) there.  

As a final remark, we would like to note that the procedure outlined in this section can be applied to derive conserved charges for the cases $\Omega_-^2 < 0$ and $\Omega_-^2 =0$. One can still construct the related $O-$matrix (\ref{Ogeneric}) using the solutions led by chiral decomposition for each case and simplify it with the initial condition (\ref{simpinitial}). Then it remains to modify the conserved charges (\ref{QCarrollPU}) of the Hamiltonian $H_{PU}$ (\ref{PUHorigin}) separately for both cases.

\section{Discussion}

In this work, we firstly interrelate two seemingly distant physical systems, \cpp  and 1-dimensional \PUo. We conclude that, via Bargmann framework, the PU oscillator is the underlying non-relativistic system of a 4-dimensional CPP gravitational wave, showing a novel and surprising aspect of the former.

Conversely, one can ED lift the PU oscillator and up in the Bargmann manifold it will become a CPP GW endowed with a Carroll symmetry. It is interesting to see such a link between Carroll group and PU oscillator which is a NR and higher derivative theory. 

Secondly, we find conserved charges for the PU oscillator originating from the symmetries of the CPP gravitational wave for any $\Omega_+ > \Omega_-$ in a novel way. Carroll boosts and Carroll translations  (\ref{CarrollB}) become $\cQ_i^{c,b}$ (\ref{QCarrollPU}). NR mass $M$ is the central extension. In addition, screw symmetry (\ref{screw}) yields the conserved energy of the PU oscillator $\cQ_s$ (\ref{screwPU}). Poisson brackets of those charges in PU phase space (\ref{QCarrollPUPoisson}) and (\ref{QCarrollHPUPoisson}) match with the ones in \cite{Andr0} for generic frequencies. Adopting the particular choice $\Omega_+ = 3 \Omega_-$, we see that our $\cQ_i^{c,b}$ (\ref{QCarrollPU}) become linear combinations of translation, boost and accelerations, namely $C^{(0)},C^{(1)}, C^{(2)},C^{(3)}$ denoted in \cite{GaMa1, AGGM} as being elements of Newton-Hooke group.
For algebraic discussions regarding $\cQ_{h1}$ (\ref{cQh1PU}) we refer a forthcoming paper \cite{Dieter}.

There are several open questions that we would like to answer in the future. For instance, we plan to investigate whether the correspondence between CPP GW and PU oscillator can be built entirely ``upstairs'', i.e., at the level of the metric. This may shed some light on the lifting of PU oscillator and link our work to that of \cite{GaMa2}. 
 
7-dimensional conformal algebra of the CPP GW allows us to derive the translation, the boost, 2 accelerations and the PU Hamiltonian together with central extension $M$ for PU oscillator with generic frequencies and that is in line with \cite{Andr0}. 
As our case is 1-dimensional, we do not have a generator of rotation. Perhaps, an interesting question would be whether one can obtain dilation and special conformal transformation for the PU oscillator via a Killing tensor of the CPP GW.

Similar arguments hold for periodic waves and ion traps \cite{IONGW}. As a next step, we would like to derive novel conserved quantities for those interesting NR systems from Carroll symmetry of the periodic waves. 

As previously mentioned, Ilderton \cite{ilderton} related CPP GW and electromagnetic vortex by classical double copy. Another interesting problem is to investigate the role of Carroll symmetry for other physical examples like vortex configuration via double copy (see e.g., \cite{Andr1}).

The maximal dimension of the conformal algebra of plane waves is 7 \cite{phonebook} and this puts a limit to our approach. Apart from the homothety which hints at a scaling property, there may be one more isometry or a conformal symmetry at most. For \PUo case, we see that screw isometry becomes the Hamiltonian. It would be interesting to work out different examples endowed with a genuine conformal symmetry \cite{Andr2}.

\vspace{1cm}

\noindent\textbf{Acknowledgements} \\
I am grateful to K. Andrzejewski, A. Galajinsky, T. Helpin, P. Horvathy, D. Van den Bleeken and U. Zorba for fruitful discussions. I am indebted  to D. Van den Bleeken also for a careful reading of the manuscript. This work was supported by the Bo\u{g}azi\c{c}i University Research Fund under grant number 21BP2.
\goodbreak



\begin{thebibliography}{99}

\bibitem{ilderton}
A.~Ilderton,
``Screw-symmetric gravitational waves: a double copy of the vortex,''
Phys. Lett. B \textbf{782} (2018), 22-27
[arXiv:1804.07290 [gr-qc]].

\bibitem{phonebook} D. Kramer, H. Stephani, M. McCallum, and E. Herlt, "Exact Solutions of Einstein’s Field Equations", 2nd ed. (Cambridge University Press, 2003) 

\bibitem{BBvort}  I.~Bialynicki-Birula,
``Particle beams guided by electromagnetic vortices: New solutions of the Lorentz, Schrodinger, Klein-Gordon and Dirac equations,''
Phys. Rev. Lett. \textbf{93} (2004), 020402
[arXiv:physics/0403078 [physics]].

\bibitem{IONGW}  P.~M.~Zhang, M.~Cariglia, C.~Duval, M.~Elbistan, G.~W.~Gibbons and P.~A.~Horvathy,
``Ion Traps and the Memory Effect for Periodic Gravitational Waves,''
Phys. Rev. D \textbf{98} (2018) no.4, 044037
[arXiv:1807.00765 [gr-qc]].

\bibitem{Paul} W. Paul, "Electromagnetic traps for charged and neutral particles", Nobel Lecture (1989), Rev. Mod. Phys. 62, 531 (1990)

\bibitem{Dehmelt} H. G. Dehmelt, "Experiments with an isolated subatomic particle at rest", Nobel Lecture (1989), Rev. Mod. Phys. 62, 525 (1990)


\bibitem{Eisenhart}  L. P. Eisenhart,  "Dynamical trajectories and geodesics", Ann. Math. 30 591–606 (1928)

\bibitem{DuvalBargmann} C.~Duval, G.~Burdet, H.~P.~Kunzle and M.~Perrin,
``Bargmann Structures and Newton-cartan Theory,''
Phys. Rev. D \textbf{31} (1985), 1841-1853

\bibitem{DGH} C.~Duval, G.~W.~Gibbons and P.~Horvathy,
``Celestial mechanics, conformal structures and gravitational waves,''
Phys. Rev. D \textbf{43} (1991), 3907-3922
[arXiv:hep-th/0512188 [hep-th]].

\bibitem{PU} A.~Pais and G.~E.~Uhlenbeck,
``On Field theories with nonlocalized action,''
Phys. Rev. \textbf{79} (1950), 145-165

\bibitem{smilga}  A.~Smilga,
``Classical and quantum dynamics of higher-derivative systems,''
Int. J. Mod. Phys. A \textbf{32} (2017) no.33, 1730025
[arXiv:1710.11538 [hep-th]].

\bibitem{Sou73}
J-M. Souriau, "Ondes et radiations gravitationnelles", 
Colloques Internationaux du CNRS No 220, p. 243. Paris (1973)

\bibitem{Carroll4GW} C.~Duval, G.~W.~Gibbons, P.~A.~Horvathy and P.~M.~Zhang,
``Carroll symmetry of plane gravitational waves,''
Class. Quant. Grav. \textbf{34} (2017) no.17, 175003
[arXiv:1702.08284 [gr-qc]].

\bibitem{LL1} J. M. Lévy-Leblond, "Une nouvelle limite non-relativiste du
group de Poincaré" Ann. Inst. Henri Poincare 3 (1965) 1

\bibitem{SenG} N. D. Sen Gupta, "On an Analogue of the Galilei Group", Il Nuovo Cimento A {\bf{44}} (1966) 512 

\bibitem{ConfCarroll} C.~Duval, G.~W.~Gibbons and P.~A.~Horvathy,
``Conformal Carroll groups and BMS symmetry,''
Class. Quant. Grav. \textbf{31} (2014), 092001
[arXiv:1402.5894 [gr-qc]].

\bibitem{Carrollvs} C.~Duval, G.~W.~Gibbons, P.~A.~Horvathy and P.~M.~Zhang,
``Carroll versus Newton and Galilei: two dual non-Einsteinian concepts of time,''
Class. Quant. Grav. \textbf{31} (2014), 085016
[arXiv:1402.0657 [gr-qc]].

\bibitem{NewCarroll} E.~Bergshoeff, J.~Gomis and G.~Longhi,
``Dynamics of Carroll Particles,''
Class. Quant. Grav. \textbf{31} (2014) no.20, 205009
[arXiv:1405.2264 [hep-th]].

\bibitem{Morand:2018tke}
K.~Morand,
``Embedding Galilean and Carrollian geometries I. Gravitational waves,''
J. Math. Phys. \textbf{61} (2020) no.8, 082502
[arXiv:1811.12681 [hep-th]].

\bibitem{Ciambelli:2019lap}
  L.~Ciambelli, R.~G.~Leigh, C.~Marteau and P.~M.~Petropoulos,
  ``Carroll Structures, Null Geometry and Conformal Isometries,''
  Phys.\ Rev.\ D {\bf 100} (2019) no.4,  046010

\bibitem{Marsot:2021tvq}
L.~Marsot,
``Planar Carrollean dynamics, and the Carroll quantum equation,''
[arXiv:2110.08489 [math-ph]].




\bibitem{homoGW}
P.~M.~Zhang, M.~Cariglia, M.~Elbistan and P.~A.~Horvathy,
``Scaling and conformal symmetries for plane gravitational waves,''
J. Math. Phys. \textbf{61} (2020) no.2, 022502
[arXiv:1905.08661 [gr-qc]].

\bibitem{GaMa1} A.~Galajinsky and I.~Masterov,
``Dynamical realizations of l-conformal Newton-Hooke group,''
Phys. Lett. B \textbf{723} (2013), 190-195
[arXiv:1303.3419 [hep-th]].

\bibitem{AGGM} K.~Andrzejewski, A.~Galajinsky, J.~Gonera and I.~Masterov,
``Conformal Newton\textendash{}Hooke symmetry of Pais\textendash{}Uhlenbeck oscillator,''
Nucl. Phys. B \textbf{885} (2014), 150-162
[arXiv:1402.1297 [hep-th]].

\bibitem{Andr0}
K.~Andrzejewski,
``Hamiltonian formalisms and symmetries of the Pais\textendash{}Uhlenbeck oscillator,''
Nucl. Phys. B \textbf{889} (2014), 333-350
[arXiv:1410.0479 [hep-th]].


\bibitem{GaMa2} A.~Galajinsky and I.~Masterov,
``Eisenhart lift for higher derivative systems,''
Phys. Lett. B \textbf{765} (2017), 86-90
[arXiv:1611.04294 [hep-th]].


\bibitem{Pl1} P.~D.~Alvarez, J.~Gomis, K.~Kamimura and M.~S.~Plyushchay,
``Anisotropic harmonic oscillator, non-commutative Landau problem and exotic Newton-Hooke symmetry,''
Phys. Lett. B \textbf{659} (2008), 906-912
[arXiv:0711.2644 [hep-th]].



\bibitem {Pl2} P.~D.~Alvarez, J.~Gomis, K.~Kamimura and M.~S.~Plyushchay,
``(2+1)D Exotic Newton-Hooke Symmetry, Duality and Projective Phase,''
Annals Phys. \textbf{322} (2007), 1556-1586
[arXiv:hep-th/0702014 [hep-th]].


\bibitem{Kohn} P.~M.~Zhang, G.~W.~Gibbons and P.~A.~Horvathy,
``Kohn's theorem and Newton-Hooke symmetry for Hill's equations,''
Phys. Rev. D \textbf{85} (2012), 045031
[arXiv:1112.4793 [hep-th]].


\bibitem{polper} P.~M.~Zhang, C.~Duval, G.~W.~Gibbons and P.~A.~Horvathy,
``Velocity Memory Effect for Polarized Gravitational Waves,''
JCAP \textbf{05} (2018), 030
[arXiv:1802.09061 [gr-qc]].

\bibitem{Brinkmann} M. W. Brinkmann, “Einstein spaces which are mapped conformally on each other,” Math. Ann. 94, 119–145 (1925).



\bibitem{zhangSL} P.~M.~Zhang, M.~Elbistan, G.~W.~Gibbons and P.~A.~Horvathy,
``Sturm\textendash{}Liouville and Carroll: at the heart of the memory effect,''
Gen. Rel. Grav. \textbf{50} (2018) no.9, 107
[arXiv:1803.09640 [gr-qc]].

\bibitem{Sarioglu:2006vc}
O.~Sarioglu and B.~Tekin,
``Topologically massive gravity as a Pais-Uhlenbeck oscillator,''
Class. Quant. Grav. \textbf{23} (2006), 7541-7550
[arXiv:gr-qc/0608085 [gr-qc]].


\bibitem{noether}
P.~M.~Zhang, M.~Elbistan, P.~A.~Horvathy and P.~Kosi\'nski,
``A generalized Noether theorem for scaling symmetry,''
Eur. Phys. J. Plus \textbf{135} (2020) no.2, 223
[arXiv:1903.05070 [math-ph]].


\bibitem{Dieter}  José Figueroa-O'Farrill, Can Gormez and Dieter Van den Bleeken, work in progress

\bibitem{Andr1} 
K.~Andrzejewski and S.~Prencel,
``From polarized gravitational waves to analytically solvable electromagnetic beams,''
Phys. Rev. D \textbf{100} (2019) no.4, 045006
[arXiv:1901.05255 [hep-th]].

\bibitem{Andr2} 
K.~Andrzejewski and S.~Prencel,
``Niederer's transformation, time-dependent oscillators and polarized gravitational waves,''
Class. Quant. Grav. \textbf{36} (2019), 155008
[arXiv:1810.06541 [gr-qc]].














\end{thebibliography}
\end{document}